\documentclass{article}

\usepackage{arxiv}

\usepackage[utf8]{inputenc} 
\usepackage[T1]{fontenc}    
\usepackage{hyperref}       
\usepackage{url}            
\usepackage{booktabs}       
\usepackage{amsfonts}       
\usepackage{microtype}      

\usepackage{amssymb}
\usepackage{amsmath}
\usepackage{mathrsfs}
\usepackage{multicol}

\usepackage{graphicx}
\usepackage{cite}

\usepackage{xcolor}
\usepackage{tabularx}

\title{Unraveling the vascular fate of deformable circulating
tumor cells via a hierarchical computational model}

\author{
Pietro Lenarda\\
 Laboratory of Nanotechnology for Precision Medicine\\
Fondazione Istituto Italiano di Tecnologia\\
Via Morego 30-16163, Genova, Italy. \\
              \texttt{pietro.lenarda@iit.it}\\
              \And
Alessandro Coclite\\
Centro di Eccellenza in Meccanica Computazionale (CEMeC)\\
Politecnico di Bari\\
Via Re David 200 – 70125 Bari, Italy.\\
              \texttt{alessandro.coclite@unibas.it}\\
              \And
Paolo Decuzzi\\
 Laboratory of Nanotechnology for Precision Medicine\\
Fondazione Istituto Italiano di Tecnologia\\
Via Morego 30-16163, Genova, Italy. \\
              \texttt{paolo.decuzzi@iit.it}\\
}

\begin{document}
\maketitle

\begin{abstract}
\textit{Introduction -- } Distant spreading of primary lesions is modulated by the vascular dynamics of circulating tumor cells (CTCs) and their ability to establish metastatic niches. While the mechanisms regulating CTC homing in specific tissues are yet to be elucidated, it is well documented that CTCs possess different size, biological properties and deformability.
\\
\textit{Methods -- } A computational model is presented to predict the vascular transport and adhesion of CTCs in whole blood. A Lattice-Boltzmann method, which is employed to solve the Navier-Stokes equation for the plasma flow, is coupled with an Immersed Boundary Method.
\\
\textit{Results -- } The vascular dynamics of a CTC is assessed in large and small microcapillaries. The CTC shear modulus ${ \rm k}_{\rm ctc}$ is varied returning CTCs that are stiffer, softer and equally deformable as compared to RBCs. In large microcapillaries, soft CTCs behave similarly to RBCs and move away from the vessel walls; whereas rigid CTCs are pushed laterally by the fast moving RBCs and interact with the vessel walls. Three adhesion behaviors are observed -- firm adhesion, rolling and crawling over the vessel walls -- depending on the CTC stiffness. On the contrary, in small microcapillaries, rigid CTCs are pushed downstream by a compact train of RBCs and cannot establish any firm interaction with the vessel walls; whereas soft CTCs are squeezed between the vessel wall and the RBC train and rapidly establish firm adhesion.
\\
\textit{Concluisons -- } These findings document the relevance of cell deformability in CTC vascular adhesion and provide insights on the mechanisms regulating metastasis formation in different vascular districts.
\end{abstract}

\section{Introduction}
\label{intro}
The shedding into the vascular network of so-called `circulating tumor cells' (CTCs) is the main mechanism
by which malignant masses colonize distant organs and tissues \cite{Nguyen2009, Wirtz2011, Joyce2009}. After leaving the primary cancerous lesion,
following a complex set of biological adaptations, CTCs face
the blood stream and, as any other blood cell, are transported away along the vascular network.
Although the mechanism by which CTCs select
their final homing tissue is not yet fully understood, experimental evidence supports
the notion that CTCs tend to more efficiently interact
with the vessel walls and eventually extravasate in microcapillaries, with diameters
in the range of a few tens of micrometers \cite{Mollica2018}.
The vascular transport, adhesion and subsequent extravasation of CTCs are regulated by local hemodynamics and biological conditions
and affected by the cell size and deformability. Individual CTCs present an average
radius ranging from $5$ to over $15 \mu$m \cite{King2015}.
More interestingly, multiple studies have documented significant
differences in deformability among cancer cells \cite{Tan2009, Sollier2014, Riahi2014}. In general, atomic force
microscopy, optical and magnetic tweezer-based assays, and micropipette aspiration studies have demonstrated that malignant cells are more
deformable than their healthy counterparts. For instance, Hrynkiewicz and his group tested multiple
cells lines using scanning force microscopy and documented a difference of about $1$ order of magnitude between healthy and cancer cells\cite{Lekka1999}. Using primary cells
from humans, oral carcinoma cells were found to be about $3$ times more compliant than healthy cells \cite{Remmerbach2009}. 
Similar observations were provided by the group of Gimzewski \cite{Cross2007}.
Recently, the group of Manalis has elegantly compared the deformability of cancer cells
(lung, breast and prostate cancer) directly to that of blood
cells (erythrocytes, leukocytes, and peripheral monocytes), using a suspended microchannel
resonator and measuring cell passage times through a
constriction \cite{Shaw2015}. The work concludes that CTC deformability can be larger,
comparable or lower than that documented for blood cells.
\\
\\
Blood flow dynamics in microvessels is primarily ruled by a non-Newtonian effect called the F\r{a}hr\ae{}us-Lindqvist effect
\cite{Zhang2009}, which is characterized
by the RBC migration away from the walls and progressive accumulation in the vessel core.
At the continuum scale, the complex rheological behaviour
of blood can be mathematically treated by adopting proper constitutive models, such as three-dimensional
neo-Hookean or viscoelastic laws \cite{Skalak1973, Kruger2011}.
However, within a microvascular network, cell-cell collisions and cell-cell adhesive interactions need to
be explicitly accounted for.
As such,
mesoscale models are required in order to analyze the spatial and temporal evolution of blood flow and its cellular
component \cite{Schiller2017}. Starting with the
pioneering work of Pozrikidis and colleagues \cite{Pozrikidis2003}, who described blood
cells using immersed boundary methods (IBM), different computational approaches
have been presented in the open literature to capture the dynamics of a multitude of deformable cells.
The group of Karniadakis
has adopted a dissipative particle dynamics method (DPD) in which the RBC membrane is represented by a coarse-grained
spring network \cite{Fedosov2010, Fedosovv2010}.
More recently, Gompper and colleagues introduced a particle-based mesoscopic simulation technique, called the smoothed dissipative particle
dynamics (SDPD) method, which combines smoothed particle hydrodynamics and dissipative particle dynamics
\cite{Gompper2014, Fedosov2014}.
\\
IBM coupled on one side with a fluid solver and on the other side with a structural solver represents a potent tool for predicting 
the transport of cells and particles in microcapillaries \cite{Peskin2002}. The structural solver serves to capture the deformation of the 
immersed object over time under different hydrodynamic stresses. This could be a network of viscoelastic springs or a membrane with a 
complex rheological behavior solved by a Finite Element procedure \cite{Saadat2018, Wing2013, Wing2014}. The latter requires heavy parallelization as compared 
to the spring network limiting the maximum number of immersed objects. 
\\
In recent years the Lattice Boltzmann method (LBM) has been widely used as a fluid solver in IBM schemes 
\cite{Falcucci2011, Succi2008, Succi2019}. In 2D problems, LBM-IBM approach with elastic spring networks has been extensively documented
in simulation of transport of RBCs, cells and particles in microcapillaries \cite{Sun2003, Coclite2017, Zhang2012}. In 3D simulations, community codes such as 
LAMMPAS (large-scale atomic/molecular massively parallel simulator) and ESPResSo (Extensible Simulation Package
for RESearch on SOft matter) have been recentely proposed \cite{Ye2018, Gekle2016}. 
\\
\\
In the literature, only a few works are focused on predicting the microvascular transport of circulating tumor cells.
For instance, Rejniak
used a $2D$ Immersed Boundary method to study the interaction of a single tumor cell with the vascular endothelium in pure 
plasma \cite{Rejniak2012}. Yan and colleagues studied the vascular adhesion of a rigid,
spherical cell in either a curved or a straight capillary, 
using a LBM method for describing blood flow, without accounting for the presence of RBCs\cite{Yan2010}. More recently, the same group adopted a DPD
computational scheme for predicting the transport of an individual tumor cell, initially released at the blood vessel wall,
in the presence of RBCs \cite{Xiao2017}. This study demonstrates that RBCs enhance CTC adhesion in small capillaries
whereas, in large vessels, CTC can
be more easily detached away
from the wall, especially at higher hematocrits. In the same work, a preliminary analysis of CTC deformability was also provided, considering a 
cell shear modulus about $25, 2.5$ and $0.25$ time larger than that of RBCs. It was concluded that softer cells can engage a larger number
of ligand-receptor bonds upon adhesion with the vascular walls. 
\\
\\
In the present work, the effect of CTC deformability on metastasis formation is analyzed using a hierarchical
computational model, where Lattice
Boltzmann and Immersed Boundary methods are combined together. The Lattice-Boltzmann method is employed
to solve the Navier-Stokes equation
governing the pure plasma flow \cite{Succi2001, Succi2008, Succi2019, Falcucci2011, en11040715}; whereas, the Immersed Boundary Method \cite{Peskin2002}
is adopted to describe the deformation and transport
of RBCs and CTCs. The cell membranes are discretized as an ensemble of linear elastic springs, connecting neighboring membrane points. Following the seminal work of Hammer and colleagues \cite{Hammer1992, King2001}, an adhesive
potential is also included to describe vascular adhesion, as mediated by the formation of individual receptor-ligand bonds treated as linear elastic
springs \cite{Hammer1992, Ye2017, Coclite2017}. The aim of this work is to elucidate the role of CTC deformability on vascular margination and subsequent adhesion in the
presence of whole blood flow.

\section{Computational method}
\label{sec:1}
The presented hierarchical computational model relies on the combination of a fluid solver for the incompressible Navier-Stokes
equation,
based on the three-dimensional D3Q19 Lattice-Boltzmann Method (LBM) \cite{Succi2001, Succi2008, Sun2003},
and a structural solver for the dynamics
of the deformable membrane, based on an Immersed Boundary Method (IBM) \cite{Peskin2002}. Details of the physical model,
governing equations and numerical
implementation are given in the sequel.

\section{Lattice Boltzmann method}
The LBM introduces a number of $N$ populations $\{ f_i \}$,$(i=0, \dots ,N-1)$ streaming along a regular lattice in discrete time steps.
These populations can be regarded as mesoscopic particles propagating and colliding. The evolution of the $N$ populations is given by
the Lattice-Boltzmann equations \cite{Succi2001}, which takes the form:
\begin{equation}
\begin{aligned}
 f_i (X+ c_i \Delta t ,t &+ \Delta t)- f_i (X,t)= \\
 &- \dfrac{\Delta t}{ \tau} [ f_i (X,t)- f^{\rm eq }_i (X,t)]+ \Delta t F_i ,
 \end{aligned}
\end{equation}
where $X$ is the spatial coordinate on a Cartesian regular lattice and $t$ is the time coordinate; $\{ c_i \}, (i=0, \dots, N-1)$
is the set of discrete velocities; $\Delta t$ is the time step; and $\tau$ is the relaxation time.
The kinematic viscosity $\nu= c^2_s (\tau- \frac{1}{2} ) \Delta t$ of the flow is related to the relaxation time $\tau$, being
$c_s = \frac{1}{\sqrt {3}} \frac{\Delta x}{\Delta t}$ the reticular speed of sound. The local equilibrium density
functions $\{ f^{\rm eq}_i \}$
are expressed by the Maxwell-Boltzmann distribution:
\begin{equation}
 f^{\rm eq}_i (X,t)=\omega_i \rho \left[ 1+ \dfrac{1}{c^2_s} \left( c_i   \cdot u \right)+ \frac{1}{2c^4_s} ( c_i \cdot u)^2 - \dfrac{1}{2c^2_s} u^2 \right],
\end{equation}
where $\{ \omega_i \}, (i=0, \dots, N-1)$ are the lattice weights, depending on the underlying lattice structure;
$\rho$ is the density and $u$ is the velocity field. 
A forcing term $f_{\rm ib}$, having the dimension of a body force density, can be incorporated via $F_i$ as:
\begin{equation}
 F_i = \omega_i (1- \dfrac{1}{2 \tau} ) \left( \dfrac{ c_i -u}{c^2_s} + \dfrac{c_i \cdot u}{c^4_s} c_i \right)  \cdot f_{ \rm ib}.
\end{equation}
It is worth noting that the fluid interacts with an immersed object only via the  forcing $f_{\rm ib}$.
In this regard, the body force density
$f_{\rm ib}$ is the term linking togheter the LB and IB modules. From equations (1-3),
macroscopic quantities can be recovered respectively
as the fluid density $\rho= \sum_i f_i $ and velocity $\rho u= \sum_i c_i f_i + \Delta t F_i /2$.
\\
On the three-dimensional square lattice with $N=19$ speeds (D3Q19) \cite{Qian1992}, the set of discrete velocities
is given by:
\begin{equation}
 c_i=\begin{cases}
      ( 0 ,0,0 )                                           ,  &  i=0  , \\
      (\pm 1,0,0 ),( 0, \pm 1,0 ),( 0,0,  \pm 1)           ,  &  i=1-6  , \\
      (\pm 1, \pm 1,0 ),(\pm 1,0,  \pm 1),( 0, \pm 1,\pm 1),  &  i=7-12  , \\
      (\pm 1, \mp 1,0 ),(\pm 1,0,  \mp 1),( 0, \pm 1 \mp 1),  &  i=13-18 
        \end{cases}
\end{equation}
with the weight, $\omega_i= 1/18$ for $i = 1, \dots ,6$, $\omega_i= 1/36$ for $i = 7, \dots ,18$, and $\omega_0= 1/3$.

\subsection{Immersed Boundary method}
\begin{figure}
 \centering
 \includegraphics[width=12.0cm]{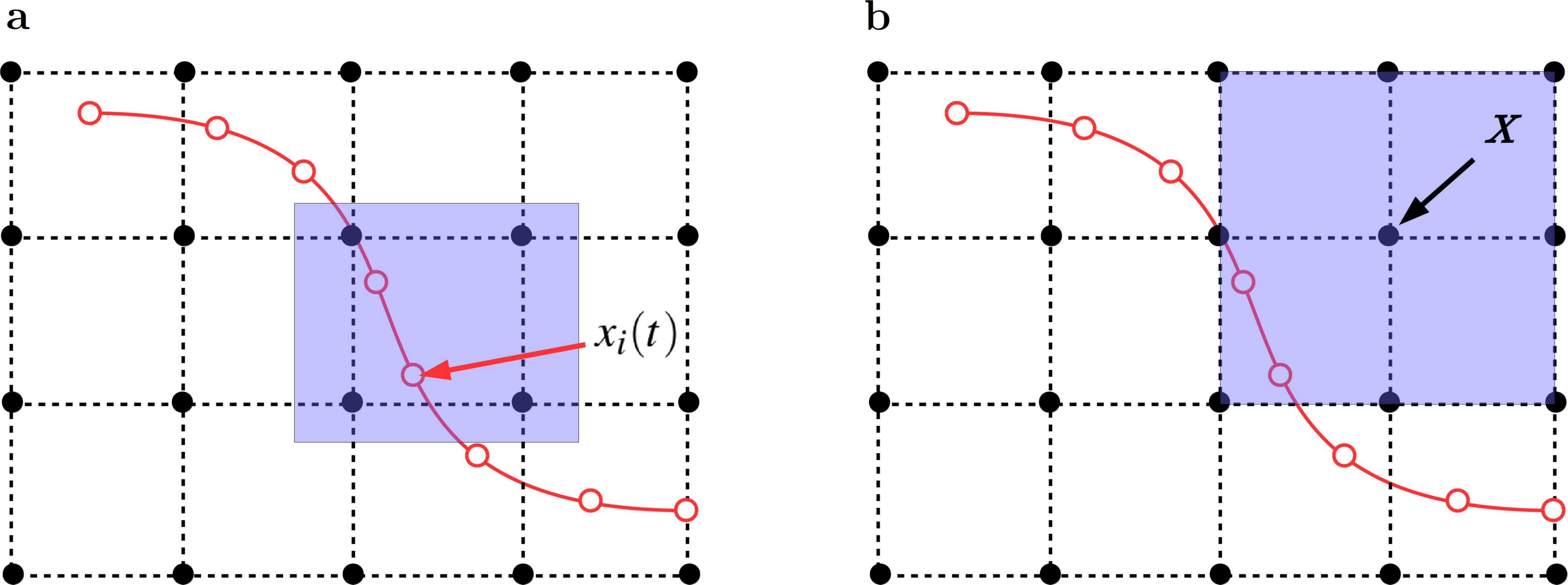}
 \caption{ \textbf{Schematic representation of the Immersed Boundary method. }
 \textbf{a}. Interpolation step: the velocity of node $x_i(t)$ is interpolated from the lattice nodes inside the interpolation stencil.
 \textbf{b}. Spreading forces: the force density acting on the fluid node $X$ is obtained from the Lagrangian nodes inside the square region.} 
 \label{fig1}
\end{figure}

In the IB module, two independent meshes are considered to
approximate respectively the immersed membranes and the fluid domain. The structure is discretized
by a moving Lagrangian mesh in which the position of each node is $x_i(t)$, while the fluid is discretized
by a fixed Eulerian mesh. The different steps in the computational algorithm are the following \cite{Peskin2002}.
First, compute the total particle forces $F_i(t)$ acting on the Lagrangian point $x_i(t)$ of the immersed object
(see \textbf{Fig.\ref{fig1}}). These forces account for the internal elastic forces, interaction forces, adhesive
forces and define the biological and mechanical behavior of the immersed object. Second, spread forces
from the Lagrangian to the Eulerian mesh via the interpolation: 
\begin{equation}
 f_{ \rm ib} \left(X , t \right) = \sum_i F_i (t) D ( x_i (t) - X  ) A_i,
\end{equation}
where the index $i$ ranges over all Lagrangian points $x_i$ inside the interpolation stencil surrounding the
Eulerian point $X$. $A_i$ is the area element associated with the Lagrangian node $x_i$. The operator $D$ is the
discretized delta function. Third, solve the Lattice-Boltzmann equation for the fluid and find the
velocity vector $u(X, t)$. Fourth, interpolate the fluid velocity to derive the velocity at each boundary
node:
\begin{equation}
 \dot{x}_i(t) = \sum_X
u(X, t) D(x_i (t) - X) ,
\end{equation}
where $X$ ranges over all Eulerian points inside the interpolation stencil surrounding $x_i$. Fifth, update
the particle position as $x_i (t + 1) = x_i (t) + \dot{x}_i(t) \Delta t$. 
\\
The choice of the discretized delta function is not unique and characterize the size of the interpolation
stencil. Let $r = x_i - X$, the Dirac delta function $D(r)$ is factorized as $D(r) = \varphi_d (r_x ) \varphi_d (r_y ) \varphi_d (r_z )$,
where $\varphi_d$ define an interpolation kernel and $d$ is the support of the stencil. Two kernel functions have
been used in the simulations:
\begin{equation*}
 \varphi_2(x)=\begin{cases}
          1- \vert x \vert    , \  & \text{if} \  0 \leq  \vert x \vert \leq 1  , \\
           0            , \  & \text{if} \  \vert x \vert >        1  
           \end{cases}   
           \end{equation*}
           and 
  \begin{equation*}         
 \varphi_4(x)=\begin{cases}
        \frac{1}{4} \left (1 + \cos ( \frac{ \pi x}{2}) \right) , \ & \text{if} \   \vert x \vert \leq 2  , \\
           0            , \  & \text{if} \  \vert x \vert >  2       
           \end{cases} 
           \end{equation*}
The choice of $\varphi_2$ leads to an interpolation stencil consisting of $2^3$ Eulerian points for the interpolations of the forces
and velocities \cite{Kruger2011, Kruger2016}, while $\varphi_4$ leads to an interpolation stencil containing $4^3$ points.

\subsection{Constitutive model for the membrane of the immersed object}
A regular mesh is used to approximate
the surface of the immersed object. This could be a spherical or biconcave capsule if it represents a
CTC or a RBC, respectively. Meridional and azimuthal angles are used to describe the surface of the
immersed object. The advantage of using a parametric description of the mesh is that the local
connectivity of each node of the network, i.e. its neighborhood structure, is simply induced by the
equations of the geometric surface. This is particularly useful when the elastic forces are computed.
Each Lagrangian node on the surface of the immersed object $x_{n,m}$ is identyfied by two parameters $0  \leq n, m \leq N$ ,
such that the total number of points on the surface is $N^2$. The stretching force $F^{ \rm s}$ acting on
each Lagrangian node $x_{n,m}$ is defined by summing up the forces on the springs connected to its
neighboring nodes $x_{n',m'} \in \{ x_{ n \pm 1,m} , x_{ n,m \pm 1 } \}$ given by \cite{Siguenza2016, Mendez2014}:
\begin{equation}
 F^{\rm s}_{n,m}= - {\rm k} \sum_ {n',m'}\dfrac{d_{n',m'} - d^0_{n,m}}{d^0_{n,m}}\dfrac{(x_{ n',m'} - x_{ n,m} )}{d_{n,m}} ,
\end{equation}
where $d_{n,m} = \vert x_{n',m'} - x_{n,m} \vert$ is the distance between node $x_{n,m}$ and its neighbor $x_{ n',m'}$ and ${\rm k}$ 
is the shear elastic
modulus.
A simple bending term can be considered as:
\begin{equation}
 F^{\rm b}_{n,m}=-{\rm k}^{\rm b} \sum_j \left( \dfrac{\theta_j -\theta^0_j}{\theta^0_j} \right) \hat{n}_j,
\end{equation}
where index $j$ runs over adjacent edges $\{ (n,m), (n \pm 1, m \pm 1) \}$, 
$\theta_j$ are the angles between adjacent triangles sourranding point $x_{n,m}$ and $\hat{n}_j$ is the outward unit vector related to edge $j$.
To enforce membrane incompressibility, a constraint on the total volume is needed:
\begin{equation}
F^{\rm v}_{n,m}
= - {\rm k}^{ \rm v} \left(1  - \dfrac{V}{V^0} \right)A_{n,m} n_{n,m} ,
\end{equation}
where ${\rm k}^{ \rm v}$ is the volume constraint factor, $V^0$ and $V$ are the volumes of the immersed object in the
reference and current configuration respectively, $n_{n,m}$ is the outward unit normal vector associated to
the Lagrangian point $x_{n,m}$ and $A_{n,m}$ is the area of the triangular element defined by the three mesh
points $\{x_{n,m} , x_{n+1,m} , x_{n,m+1} \}$. To evaluate the normal unit vector and the area $A_{n,m}$ consider $\xi_1 = x_{n+1,m} -
x_{n,m}$ and $\xi_2 = x_{n,m+1} - x_{n,m}$,
then the unit outward normal vector is $n_{n,m} = \xi_1 \wedge \xi_2 / \vert \xi_1 \wedge \xi_2 \vert$ and the area of
the triangular element is $A_{n,m} = \vert \xi_1 \wedge \xi_2 \vert / 2$. The volume is computed at each time iteration via the
discrete Green theorem as $V = \sum_T n_T \cdot c_T A_T / 3$, where the summation index
runs over all triangles $T=\{ x_{n,m} , x_{n+1,m} , x_{n,m+1} \}$, $n_T$ the unit normal vector to the surface
associated to triangle $T$, $A_T$ is the area of
the triangular element, and $c_T = (x_{n,m} + x_{n+1,m} + x_{n,m+1} ) /3$ is the baricenter of the triangle $T$.
Another constraint is imposed on the conservation of the total surface of the cell:
\begin{equation}
F^{\rm a }_{n,m}
= -{\rm k}^{ \rm a} \left(1 - \dfrac{A}{A_0} \right) \ \zeta_{n,m} ,
\end{equation}
where ${\rm k}^{ \rm area}$ is the area constraint factor and $\zeta_{n,m}$ is the unit vector pointing from the centroid of
triangle $T =\{x_{n,m} , x_{n+1,m} , x_{n,m+1 } \}$ to the vertex $x_{n,m}$ . This constraint is applied in the case of RBCs and
CTCs to account for their cytoskeleton \cite{Suresh2005, Mills2004}.

\subsection{Particle-particle interactions}
The cell-cell interactions can be approximated with the Morse
potential given by \cite{Zhang2009}:
\begin{equation}
 U_M(r) = D_e (e^{2 \beta (r_0 - r)} - 2e^{ \beta(r_0 - r)} ) ,
\end{equation}
where $r$ is the distance between two cells and $D_e$ is the energy well depth. 
Note that the calibration of the Morse potential
can be done following the same 
procedure in \cite{Ye2017} returning $D_e=0.04 k_B T$, where $k_B T=4.14 \times 10^{-21} N \cdot m$.
Let $d_{n,m}=\vert x_{n,m} - x_{n',m'} \vert$
be the distance between two nodes on different structures.
The interaction force between node $x_{n,m}$ on the surface of the
current cell and node $x_{n',m'}$ on another cell is given by:
\begin{equation}
 F^{\rm inte}_{n,m}=\begin{cases}
          -2 \beta D_e U_M \dfrac{x_{n,m}- x_{n',m'}}{\vert x_{n,m}- x_{n',m'} \vert} , & \text{if} \ d_{n,m} < d_{ \rm cut}  \\
                      0 ,  & \text{otherwise}.
                    \end{cases}
                    \end{equation}
The Morse interaction potential is implemented between two nodes of separate cells if they are within
a cutoff distance $d_{ \rm cut}$. This type of interaction consists of a high short-range repulsive force when $r <
r_0$ and a low long-range attractive force for $r > r_0$. Parameters used are: $\beta = 3.84 \ \mu$m$^{-1}$, $r_0 = 0.5 \ \mu$m,
and $d_{ \rm cut} = 1.5 \ \mu$m \cite{Zhang2009, McWhirter2009}.

\subsection{Wall-particle interactions}
Ligand and receptor molecules are distributed over the cell and blood
vessel surfaces, respectively. Ligand molecules are modeled as linear springs which tend to
establish bonds with receptors on the vascular wall. Let $x_{n,m}$ be a point on the particle surface, $x_{\rm wall}$ be
the normal projection of $x_{n,m}$ on the wall of the channel, and $l_{n,m} = \vert x_{n,m} - x_{\rm wall} \vert$ 
be the distance between
a point on the cell surface and the corresponding point on the wall, when the distance between the
surface of the capsule and the wall is less than a critical distance $d_{ \rm wall}$, the adhesive force acting on $x_{n,m}$ is
defined as \cite{Rejniak2012, Xiao2017}:
\begin{equation}
  F^{\rm adh}_{n,m}=\begin{cases}
          - \sigma(l_{n,m} - \lambda) \dfrac{ x_{n,m}- x_{\rm wall}}{ l_{n,m}} , \ & \text{if} \ d^{ \rm w}_{n,m} < d_{ \rm wall} \\
                      0 , \ & \text{otherwise}.
                    \end{cases}
\end{equation}
where $d^{\rm w}_{n,m}=d( x_{n,m}, \rm wall)$ is the distance of $x_{n,m}$ from the wall, $\sigma$ is the
adhesion constant, $\lambda$ is the equilibrium separation distance
for the spring and $d_{\rm wall} = 2 \lambda$ is
the critical distance.
It is here important to recall that the mathematical model for cell-wall adhesion mediated by receptor-ligand interactions was originally presented by Hammer and colleagues and applied to study the rolling dynamics of leukocytes on endothelial surfaces \cite{Hammer1992}. Since then, this Adhesive-Dynamics model (AD) has been extended  to study many biologically relevant problems, such as the hydrodynamic recruitment of rolling
leukocytes \cite{King2001}, platelet-surface and platlet-platlet interactions \cite{Mody2005, Wang2013}. 
Note also that the current model does not account for the stochastic formation and rupture of ligand receptor bonds, which can be readily included following previous works by the authors and others \cite{Hammer1992, King2001, Coclite2017}.

\section{Results and discussion}

\begin{figure}[h!]
 \centering
 \includegraphics[width=12cm]{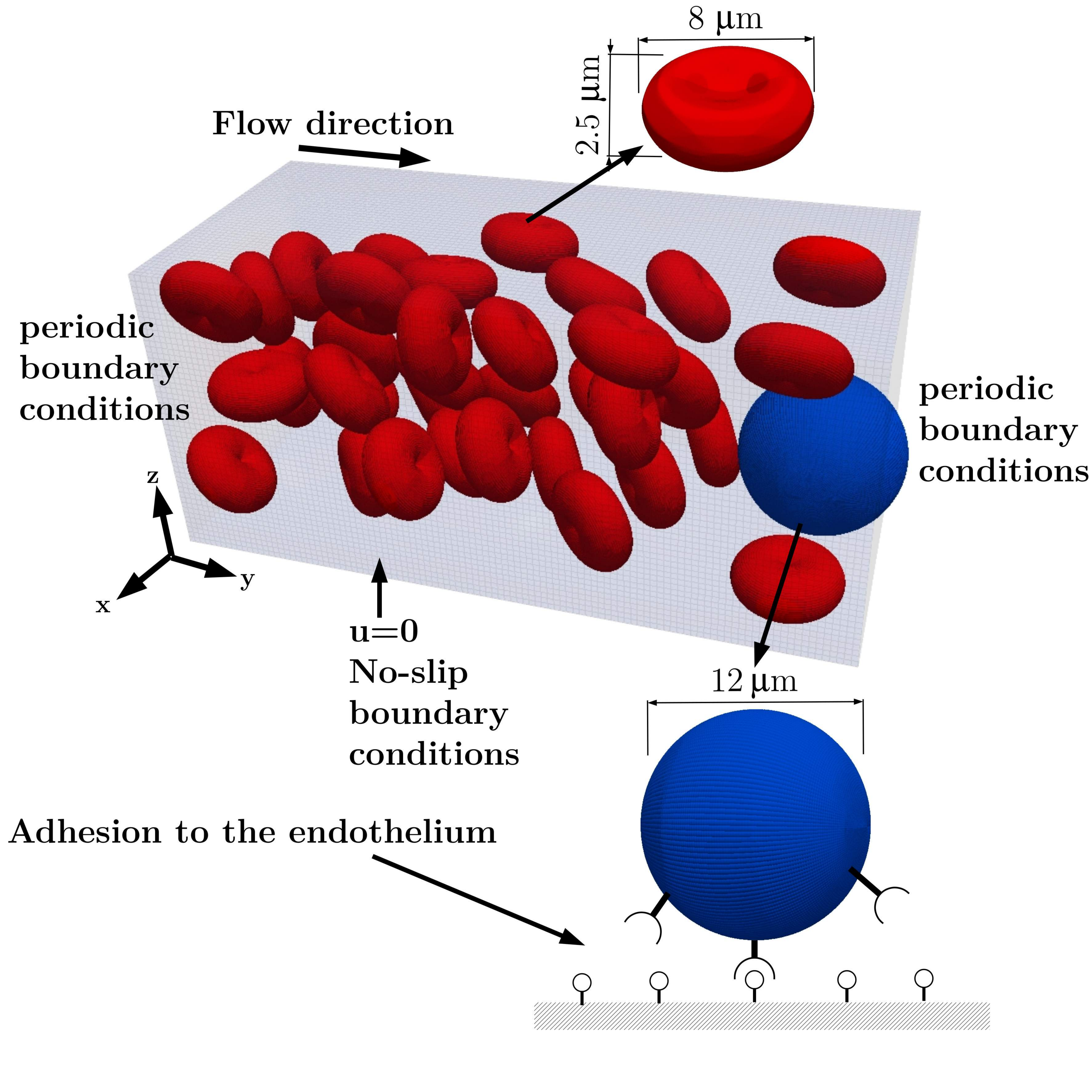}
 \caption{ \textbf{Schematic representation of a CTC in whole blood flow.} 
 A spherical, deformable
circulating tumor cell (CTC) is transported downstream in a whole blood flow. The CTC interaction
with the vessel wall is mediated by the formation of receptor-ligand bonds.}
 \label{fig2}
\end{figure}

The proposed hierarchical computational model combines a Lattice-Boltzmann (LBM) algorithm, for
solving the fluid flow, with an Immersed Boundary method (IBM), for determining particle-fluid and
particle-wall interactions. As such, the computational model can efficiently deal with multiple scales 
and different biophysical problems, spanning from the ligand-receptor adhesive interactions
(molecular scale) to the deformation of cell membranes (mesoscopic scale) and the transport of
multiple red blood cells (RBCs) in a capillary flow (macroscopic scale). This is schematically
presented in \textbf{Fig.\ref{fig2}}. The computational model is first validated against known test cases: a
deformable spherical capsule in a linear shear flow; the stretching of a red blood cell under uniaxial
loading. 
Finally, the model is applied to document the vascular transport and
adhesion dynamics of a single circulating tumor cells (CTCs), in whole blood flow, in microvessels of
different calibers.
It is worth noting that in the present paper, only one CTC is considered.
Given the low abundance of CTCs in blood, this condition is indeed physiologically sound. 
Other blood cells, such as leukocytes, platelets, monocytes and so on, are not explicitly modeled in this problem in that they are far less abundant than RBCs. Indeed, RBCs represent up to $95 \%$ of all cellular components of blood and are responsible of the peculiar blood rheology. In small vessels, ranging from $10$ to $40 \; \mu \rm m$, the volume fraction of RBCs varies from $15$ to about $35 \%$ \cite{Fedosov2010}. In the present work, the hematocrit has been fixed to the average value of $20 \%$. Indeed, higher values for the hematocrit are associated with higher computational burden. 
The fluid solver used in this model relies on the Lattice-Boltzmann method which is very accurate and effective in terms of single cost per time iteration. However, a Lattice-Boltzmann solver is explicit in time and, as such, requires small time steps (in the order of $10^{-7}$ sec) to ensure stability. The effect of the mesh resolution on the accuracy of the solution is provided for the first test case.

\subsection{A deformable spherical capsule in a linear shear flow: Test case 1}

\begin{figure}[h!]
 \centering
 \includegraphics[width=12cm]{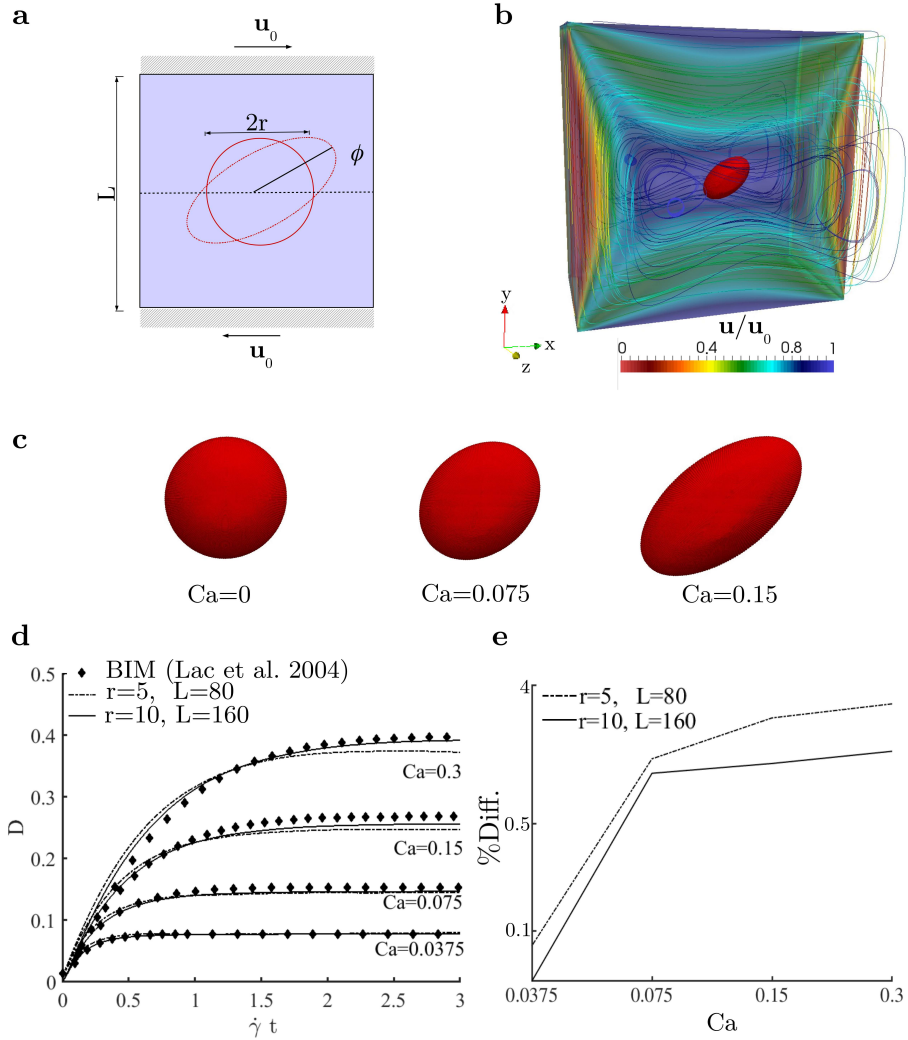} 
 \caption{ \textbf{A deformable spherical capsule in a linear shear flow.} \textbf{a}.
 Schematic representation of the problem and computational domain.
 \textbf{b}. Normalized velocity field at steady state (${\rm Ca}=0.075$).
 \textbf{c}. Steady state configurations of spherical capsules for ${\rm Ca} =0, 0.075$, and  $0.15$.
 \textbf{d}. Variation of the Taylor number $D$ with time, for ${\rm Ca} = 0.0375, 0075, 0.15,$ and $0.3$ 
 (dashed line: low resolution; solid line: high resolution; solid dots: BIM simulations by Lac et al. \cite{Lac2004}).
 \textbf{e, f}. Percentage difference between BIM simulations and the present solution, at steady state,
 for different ${\rm Ca}$ and mesh resolutions (low resolution: $r= 5$, $L=80$; high resolution $r= 10$, $L=160$).}
 \label{fig3}
\end{figure}

A spherical capsule of radius $r$ is
placed in the center of a cubic box of size $L$ and subjected to a linear shear flow, realized by moving
the top and bottom walls with velocities $u_0$ and $-u_0$, respectively (\textbf{Fig.\ref{fig3}a}). The size of the fluid
mesh is $L = 16r$. $6400$ Lagrangian points are used to discretize the capsule membrane. The fluid is
considered as quasi-stationary in the so-called Stokes regime with a Reynolds number $Re = {10}^{-2}$ and a
shear rate $\dot{\gamma}=\nu Re / r^2$, where $ \nu = 1 / 6$ is the lattice viscosity. Thus, the magnitude of the imposed
velocity $u_0$ is given as $u_0 = \dot{\gamma} L / 2$. The capsule has a shear elastic modulus $k$. The capillary number ${\rm Ca}$
$({\rm Ca} = \dot{\gamma} \nu r / k )$, which represents the relative effect of the viscous drag over surface tension, provides
a measure of the capsule deformability: ${\rm Ca}=0$ means a rigid capsule ($k \rightarrow + \infty$); ${\rm Ca}>0$ means a
deformable capsule.
\\
In a linear shear flow, the capsule rotates and deforms assuming eventually the shape of an ellipsoid
(\textbf{Fig.\ref{fig3}b,c}), with $r_a$ and $r_b$ being the two principal axes of the ellipsoid. \textbf{Fig.\ref{fig3}b} gives the steady
state, normalized velocity distribution $u / u_0$ within the box, for ${\rm Ca}=0.075$. Four recirculation areas are
clearly developing around the elongated capsule. The deformation of the capsule is quantified via the
Taylor parameter $D$ ($D = (r_a - r_b ) / (r_a + r_b )$). Deformed configurations of the capsule at steady state
are shown in \textbf{Fig.\ref{fig3}c} for three representative values of the capillary number, namely ${\rm Ca}=0, 0.075$ and $0.15$.
Indeed, the larger is the capillary number, the larger is the capsule deformation $D$. This is
also presented quantitatively in \textbf{Fig.\ref{fig3}d}, where the Taylors number $D$ is plotted versus the non-dimensional time
$\dot{\gamma} t$, for different values of the capillary number ${\rm Ca} (= 0.0375, 0.075, 0.15$ and $0.3)$.
$D$ increases with time until a steady state configuration is reached for $\dot{\gamma}t \geq 3$. For longer times, the
capsule membrane rotates along its own shape with a inclination angle $\phi$,
as predicted by the known tank-treading condition \cite{Lac2004}. In
\textbf{Fig.\ref{fig3}d}, the present solution is compared with data from a neo-hookean membrane model, solved using
the Boundary Integral method (BIM), by
Lac and colleagues \cite{Lac2004} for two values of the mesh resolution, namely low ($r=5, L=80$, dotted lines)
and high ($r=10, L=160$, solid lines) resolution. The presented results are in good agreement with the
BIM data, for different ${\rm Ca}$ values \cite{Lac2004}. Notice that the agreement between the two solutions improves
as the mesh resolution increases. This appears to be particularly relevant at large ${\rm Ca}$, which
corresponds to more deformable capsule. The difference between the two numerical solutions is
plotted as a function of ${\rm Ca}$ in \textbf{Fig.\ref{fig3}e}, at steady state. This difference grows with the capillary
number ${\rm Ca}$, as previously pointed out \cite{Mendez2014}, but it stays well below $4 \%$ for all considered cases.
Table.\ref{Table_D} summarizes the $D$ values for different ${\rm Ca}$ numbers obtained by BIM simulations
and the current
method, demonstrating that the difference ranges from $0.047\%$ for ${\rm Ca}=0.0375$ to $3.04\%$ for ${\rm Ca}=0.3$. 
In Table.\ref{Table_phi}, are reported 
the values of the capsule angle $\phi/\pi$ for ${\rm Ca}=0.0375, 0.075$ and $0.15$. The angle $\phi$ is computed as\\ 
$\phi=0.5 \times \arctan \left(2 r_a r_b / (r^2_b-r^2_a)\right)$.

\begin{table*}[ht!] 
\centering 
\caption{Evolution of the Taylor parameter $\rm D$ as a function of the capillary number
$ \rm Ca$.} 
\begin{tabular}{c | c | c  |c | c | c  }
\hline  \noalign{\smallskip} \hline\noalign{\smallskip} 
\multicolumn{1}{c}{{\rm Ca}} &
\multicolumn{1}{c}{$\rm D$ Lac (2004) } &
\multicolumn{1}{c}{$\rm D$ $(r=10, L=160)$ } &
\multicolumn{1}{c}{$\%$ Diff.} &
\multicolumn{1}{c}{$\rm D$ $(r=5, L=80)$ } &
\multicolumn{1}{c}{$\%$ Diff.} 

\smallskip
\\
\hline 
 0.0375   & 0.0798 & 0.0794 & 0.04 & 0.0790 & 0.080 \\
 0.075    & 0.1594 & 0.1487 & 1.07 & 0.1461 & 1.33   \\
 0.15     & 0.2718 & 0.2594 & 1.24 & 0.2491 & 2.49   \\
 0.3      & 0.4053 & 0.3904 & 1.49 & 0.3749 & 3.04    \\

\hline  \noalign{\smallskip}\hline\noalign{\smallskip}
\end{tabular}\label{Table_D}
\end{table*}

\begin{table*}[ht!] 
\centering 
\caption{Evolution of the inclination angle $\phi / \pi$ as a function of the capillary number
$ \rm Ca$.} 
\begin{tabular}{c | c | c  |c  }
\hline  \noalign{\smallskip} \hline\noalign{\smallskip} 
\multicolumn{1}{c}{{\rm Ca}} &
\multicolumn{1}{c}{$\phi / \pi$ Lac (2004) } &
\multicolumn{1}{c}{$\phi / \pi$ $(r=5, L=80)$ } &
\multicolumn{1}{c}{$\%$ Diff.} 

\smallskip
\\
\hline 
 0.0375   & 0.21739 & 0.21929 & 0.19\\
 0.075    & 0.18391 & 0.19694 & 1.3030    \\
 0.15     & 0.15826 & 0.16401 & 0.5750  \\

\hline  \noalign{\smallskip}\hline\noalign{\smallskip}
\end{tabular}\label{Table_phi}
\end{table*}

\subsection{The stretching of a red blood cell under uniaxial loading: Test case 2}
\begin{figure}
 \centering
 \includegraphics[width=12cm]{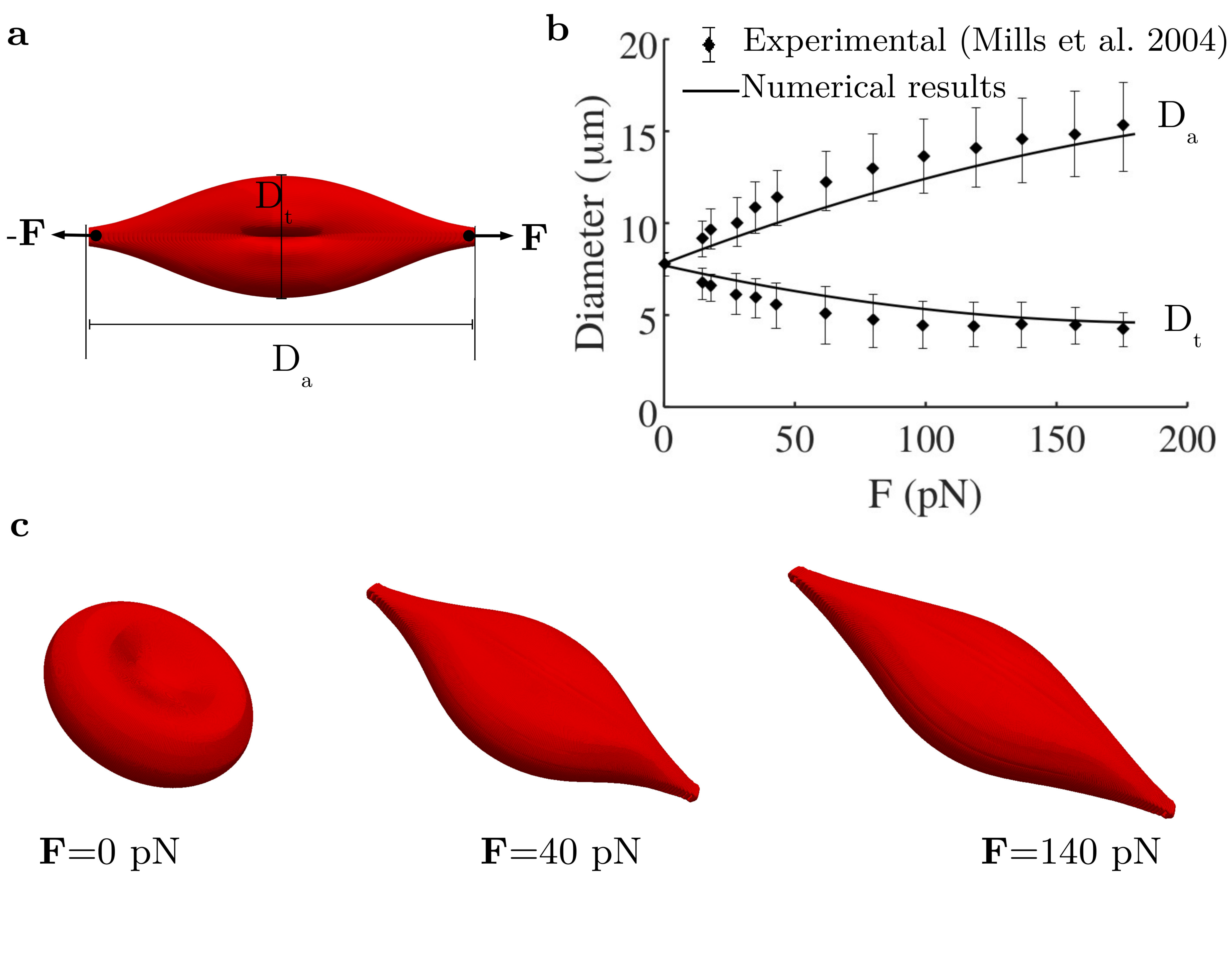}
 \caption{ \textbf{The stretching of a red blood cell under uniaxial loading.}
\textbf{a}. Schematic representation
of the problem. Pulling force $\rm F$ is applied at two opposite sides of the RBC membrane. 
\textbf{ b}. The variation of the axial and transverse diameters (${\rm D}_a$, ${\rm D}_t$) at steady state, for different
pulling forces $\rm F$. (solid dots: experimental results from Mills et al. \cite{Mills2004}; solid lines: present
hierarchical model). 
\textbf{c}. Steady state configurations of a RBC subjected to an uniaxial pulling force $\rm F$ of
$0, 40$ and $140$ pN.
 }
 \label{fig4}
\end{figure}

A single red blood cell (RBC)
is stretched longitudinally by applying a force $\rm F$ at two opposite sites of the cell membrane (\textbf{Fig.\ref{fig4}a}).
After a transient phase, the elastic reaction force ${\rm F}^{\rm el}$ arising at the cell membrane balances out the
external applied force $\rm F$ so that a steady deformation is achieved. This case serves to predict the
stretching of a RBC in a pulling test realized using an optical tweezer \cite{Suresh2005, Mills2004}. Briefly, two silica
microbeads are attached at opposite sites of the cell membrane: one bead is anchored to the surface
of a glass slide, while the other one is trapped by a laser beam. By moving the bead attached to the
glass slide, a well defined strain is applied to the cell. At equilibrium, the diameter ${\rm D}_a$ in the pulling
direction (axial) and the diameter ${\rm D}_t$ orthogonal to the pulling direction (transverse) are measured for
each value of the applied force. This is documented by solid dots in \textbf{Fig.\ref{fig4}b}. Experimentally the 
traction force $\rm F$ ranges from $0$ to $180$ pN.
\\
The $3D$ shape of the RBC is given by the following parametric equations, for $0 \leq \theta \leq 2 \pi, 0 \leq \phi \leq \pi$:
\begin{equation}
\begin{cases}
 x=    a \alpha \sin(\theta) \cos(\phi)         \\
 y=     \dfrac{a}{2} (0.207 + 2.003 \sin^2 (\theta) - 1.123 \sin^4 (\theta)) \cos(\theta)         \\
 z=     a \alpha \sin(\theta) \sin(\phi)
              \end{cases}
\end{equation}
where $a=0.74 {\rm r}_{\rm rbc}$, with ${\rm r}_{\rm rbc}=3.91 \ \mu$m being the equivalent RBC radius and $\alpha=1.39$. 
Note that this parametric equation is equivalent to the well known Evans-Fung formula in cartesian coordinates
\cite{Suresh2005, Sui2007, Poz2003}.
The RBC shear
modulus is taken as ${\rm k}_{\rm rbc}=8.3 \ \mu$N$/$m \cite{Mills2004}.
The solid lines in \textbf{Fig.\ref{fig4}b} report the values
of ${\rm D}_a$ and ${\rm D}_t$ computed at equilibrium for different values of $\rm F$ via the present hierarchical model. As
expected, the diameter ${\rm D}_a$ increases while ${\rm D}_t$ decreases with the applied force $\rm F$. The computed axial and
transverse diameters ${\rm D}_a$ and ${\rm D}_t$ are in good agreement with the experimental data \cite{Mills2004}. 
\textbf{Fig.\ref{fig4}c} shows
representative equilibrium configurations of the RBC under different applied forces. The proposed
hierarchical model captures correctly the mechanical deformation of RBCs with an overall error lower
than $10\%$. Values of the diameters $D_a$ and $D_t$ for different applied forces $F$ are reported 
in Table \ref{Table_diam} and compared with experimental data by \cite{Mills2004}. The RBC bending modulus ${\rm k}^b_{\rm rbc}$ and 
volume constraint ${\rm k}^v_{\rm rbc}$ can be estimated as previously documented in \cite{Fedosov2010, Mills2004}, returning
${\rm k}^b_{\rm rbc}=2.4 \times 10^{-19}$ J ($=0.0016$ in LB units) and ${\rm k}^v_{\rm rbc}=10$ (LB units).

\begin{table*}[ht!] 
\centering 
\caption{Measured axial and transversal RBC diameters under a stretching test.} 
\begin{tabular}{c | c | c  |c |c  }
\hline  \noalign{\smallskip} \hline\noalign{\smallskip} 
\multicolumn{1}{c}{$\rm F$(pN)} &
\multicolumn{1}{c}{${\rm D}_a$($\mu$m) Mills (2004) } &
\multicolumn{1}{c}{${\rm D}_a$($\mu$m)} &
\multicolumn{1}{c}{${\rm D}_t$($\mu$m) Mills (2004) } &
\multicolumn{1}{c}{${\rm D}_t$($\mu$m)} 
\smallskip
\\
\hline 
 20 &9.65672 $\pm$ 1.08955    & 9.19057 & 6.61194 $\pm$ 0.73134  & 6.31599 \\
 40   &11.41791 $\pm$ 1.50746 & 9.95102  & 5.58209 $\pm$ 1.22388 & 6.08232    \\
  60   &12.23881 $\pm$ 1.61194 & 10.98340  & 5.09851 $\pm$ 1.56716 & 5.81418  \\
100  & 13.64179 $\pm$ 2.01309 & 12.19840 & 4.44179 $\pm$ 1.26866 & 5.09710  \\
120 & 14.08955 $\pm$ 2.14925 & 12.69840 & 4.40299 $\pm$ 1.20896 &  4.70220 \\
140 & 14.58209 $\pm$ 2.31343 & 13.77490 & 4.50746 $\pm$ 1.22388 &  4.52800 \\
160 & 14.83582 $\pm$ 2.32836 & 14.58190 &4.46269 $\pm$ 1.01256&  4.30200 \\
180 & 15.34328 $\pm$ 2.41791 & 15.2200 &4.25373 $\pm$ 0.94030 &  4.14313 \\
\hline  \noalign{\smallskip}\hline\noalign{\smallskip}
\end{tabular}\label{Table_diam}
\end{table*}

\subsection{Margination dynamics of circulating tumor cells with different deformability}
Cancer spreading to
distant tissues (metastasis) involves the shedding in the circulation of malignant cells from a primary
lesion; their vascular transport, adhesion, extravasation and proliferation \cite{Peer2007, Wirtz2011, Joyce2009, Mollica2018}.
During their journey, CTCs reach peripheral vascular beds, with blood vessels characterized by a
diameter ranging between $10$ and $40 \ \mu$m. In these small vessels, the presence of RBCs would favor
the margination of CTCs towards the vascular endothelium, just like for leukocytes in an inflamed
vessel. While the shear modulus of a RBC is in the range of ${\rm k}_{\rm rbc} = 5 - 10 \ \mu$N $/$m \cite{Suresh2005}, CTC
deformability can vary from fractions of $1$ kPa (soft) up to $100$ kPa, which is $20$ times higher than a
RBC (rigid), under physiological conditions \cite{Guz2014, Shaw2015}. In this section, the effect of cell deformability on
CTC vascular dynamics is analyzed.
\\
The proposed hierarchical computational
model is applied to study the transport of an ensemble of RBCs and a single CTC within a capillary for an hematocrit
of $20 \%$, which is a physiological value in the microcirculation. It is well known that, due to their
deformability, RBCs move away from the walls and migrate to the center of the capillary \cite{Zhang2009, Fedosovv2010, Maeda1996}.
This results in the formation of a region depleted of cells next to the wall, which is called the cell-free layer
(CFL). This non-Newtonian effect known as the F\r{a}hr\ae{}us-Lindqvist effect \cite{Fedosovv2010} is
responsible for the modulation of the blood viscosity.

Following \cite{Ye2017, Ye2018}, a capillary with a square cross section of $H=W$ and lenght $L$ is considered.
The
equivalent radius of the RBC is ${\rm r}_{rbc} = 3.9 \ \mu$m, so that $V_{\rm rbc} = 94.1 \mu$m$^3$. Following the same equation
described in the previous section, RBCs are modelled as biconcave membranes.
\\
A number $N_{\rm rbc}$ of vectors, representing the RBC and CTC centers of mass, are uniformely generated in space
and positioned in the volume $V_{\rm cap}$. Each RBC has initially a random orientation with respect to the 
$x$- and $z$- axes. For each RBC, the number of Lagrangian points is $4900$. Periodic boundary conditions
are imposed in the flow direction at the inlet and outlet sections of the capillary, while no-slip velocity
boundary conditions are prescribed on the remaining walls. Bounce-back boundary conditions are
employed to treat the no-slip velocity conditions at the walls. The Reynolds number is fixed to be $Re =
2.5 \times  {10}^{-2}$. The blood flow is driven by a constant body force density $\rho f$,
which is equivalent to prescribe
a pressure difference over the lenght of the capillary given by $\Delta P /L = 16 u_0 \nu \rho / D^2$,
where $u_0 = \nu Re /H$
is the peak velocity in the flow direction, $D = 4 HW / 2 (H + W) = H$ is the hydraulic diameter of the
channel. The non-dimensional time $\dot{\gamma} t$ is considered, where $\dot{\gamma} = 4 u_0 /H$ is the shear rate. The capillary
number is defined as ${\rm Ca} = \nu \dot{\gamma}{\rm r}_{\rm rbc} / {\rm k}_{\rm rbc}$ and fixed to $0.026$ for the RBCs.
The Lattice resolution is $\Delta x =
0.5714 \ \mu$m and
the Lattice-Boltzmann viscosity is given by $\nu=\frac{1}{6} \Delta x^2 {\Delta t}^{-1}$. Note that, in dimensional
units, the viscosity is equal to $\nu=1.2\times {10}^{-6}$ m$^2$s$^{-1}$ and the plasma density is $\rho = 1000$ Kg$/$m$^3$.
A capillary with a square cross section of $H=W=25 \ \mu$m and lenght $L=60 \ \mu$m is considered.
The CTC is modelled as a spherical capsule having a radius ${\rm r}_{ \rm ctc} = 12 \ \mu$m and discretized with $6400$
Lagrangian points. A Poiseuille flow with Reynolds number $Re = 0.25 \times {10}^{-2}$ is assumed. Periodic
boundary conditions are prescribed at the front and back walls of the capillary, while no-slip velocity
boundary conditions are imposed on the remaining walls. Table\ref{Table_param} collects the values of all physical
parameters used in this simulation. The initial position of the CTC center of mass is $(x_0 , y_0 , z_0 ) =
(W / 2 , L - 2{\rm r}_{ \rm ctc} , H / 2 - {\rm r}_{ \rm ctc} )$.
\begin{table*}[ht!] 
\centering 
\caption{Table listing the parameters used in the simulations.} 
\begin{tabular}{c | c | c }
\hline  \noalign{\smallskip} \hline\noalign{\smallskip} 
\multicolumn{1}{c}{Parameters} &
\multicolumn{1}{c}{Symbol} &
\multicolumn{1}{c}{Value}
\smallskip
\\
\hline 
Tube height, size    & $H=W, D$ & $25, 15 \ \mu$m \\
Tube segment lenght      & $L$ & $60 \ \mu$m \\
RBC radius     & ${\rm r}_{\rm rbc}$ & $4 \ \mu$m \\
CTC radius     & ${\rm r}_{\rm ctc}$ & $12 \ \mu$m \\
RBC count     & $N_{\rm rbc}$ & Variable \\
Hematocrit     & ${\rm H_t} $ & $ N_{\rm rbc} V_{\rm rbc} / V_{\rm cap} $ \\
Plasma kinetic viscosity     & $\nu$ & $1.2 \times 10^{-6}$ m$^2$/s \\
RBC stiffness modulus    & ${\rm k}_{\rm rbc} $ & $10  \mu$ N/m \\
RBC bending modulus    & ${\rm k}^b_{\rm ctc} $ & $2.4 \times 10^{-19}$ J \\
RBC volume constraint  & ${\rm k}^v_{\rm ctc}$ & $10$  \\
CTC stiffness modulus    & ${\rm k}_{\rm ctc}$ & $5-200  \mu$ N/m \\
Reynolds number    & $Re$ & $2.5 \times 10^-2$ \\
Center velocity (no cells)     & $u_0$ & $\nu Re /H $ \\
Shear rate   & $ \dot{\gamma} $ &  $4 u_0 /H $ \\
Pressure gradient   &  $f=\Delta p /L$ & $16 \nu \rho u_0 /H^2$ \\
Lattice resolution  & $ \Delta x $ & $0.5714 \ \mu$m \\
Time step & $ \Delta t $ & $ \Delta x^2/6 \nu $ \\
Capillary number & ${\rm Ca} $ & $\nu \dot{\gamma} {\rm r}_{\rm rbc}/ {\rm k}_{\rm rbc} $ \\
Adhesive number & $ Ad $ & $\sigma / \nu \dot{\gamma} {\rm r}_{\rm rbc } $ \\
Receptor-ligand resting lenght& $ \lambda $ & $ 50$ nm \\

\hline  \noalign{\smallskip}\hline\noalign{\smallskip}
\end{tabular}\label{Table_param}
\end{table*}

\begin{figure}[h!]
 \centering
 \includegraphics[width=12cm]{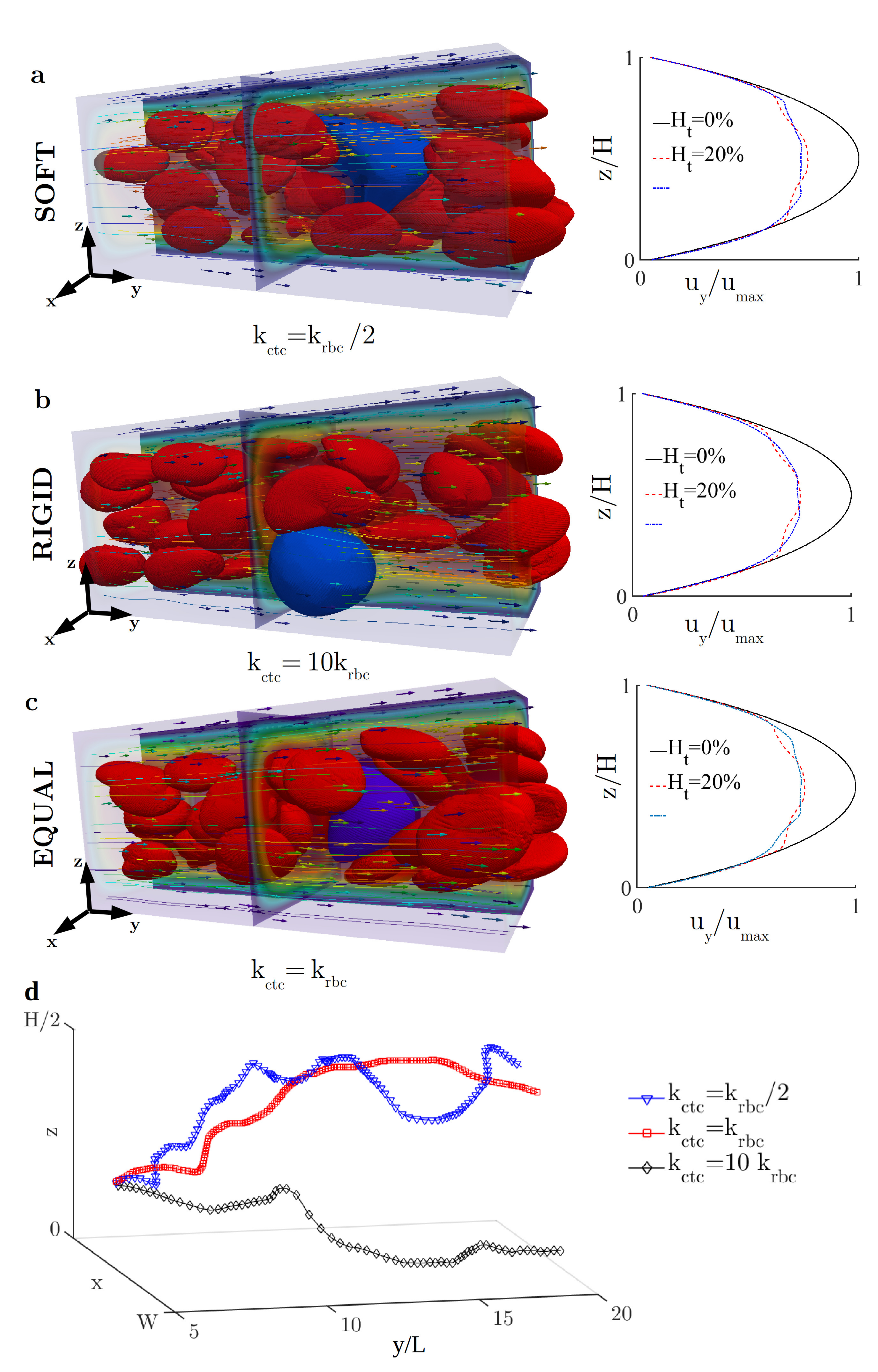}
 \caption{ \textbf{Margination dynamics of a CTC with different deformability in whole blood flow.} \textbf{a-c}. 
 Representative images of RBC and CTC distribution within a capillary whole blood flow (${\rm H_t}
=20 \%, D=25 \ \mu$m) for the `rigid' (\textbf{a}), `soft' (\textbf{b}) and `equal' (\textbf{c}) cases. The right columns show the
velocity profile compared to the
pure plasma parabolic profile. \textbf{d}. Trajectories of the center of mass of the CTC for
${\rm k}_{ctc} = \{ {\rm k}_{rbc} /2, {\rm k}_{rbc}, 10 {\rm k}_{rbc} \}$ (black line:
${\rm k}_{ctc} =10 {\rm k}_{rbc}$; blue line:
${\rm k}_{ctc} ={\rm k}_{rbc} /2$ ; red line:
${\rm k}_{ctc} ={\rm k}_{rbc}$).
 }
 \label{fig5}
\end{figure}

Three different CTC deformability values are considered: a CTC softer
than RBCs (${\rm k}_{\rm ctc} = 0.5 {\rm k}_{\rm rbc}$: SOFT); a CTC stiffer than RBCs (${\rm k}_{\rm ctc} = 10 {\rm k}_{\rm rbc}$: RIGID); a CTC as
deformable as a RBC (${\rm k}_{\rm ctc} = {\rm k}_{\rm rbc}$: EQUAL). \textbf{Fig.\ref{fig5}a,c} show the RBC distribution and the CTC
location within a longitudinal section of the capillary, at time $\dot{\gamma}t=100$, for the `soft', `rigid' and `equal'
cases. On the right, fluid velocity profiles are shown. Specifically, the different velocity profiles are referred to the
classical Poiseuille case in the absence of RBCs (black line); the time-averaged
velocity profile for ${\rm H_t} =20\%$ with RBCs and no CTC (red line); and the time-averaged velocity profile
for ${\rm H_t} =20 \%$ with RBCs and the CTC (blue line). The presence of the CTC does not change
significantly the time-averaged velocity profile, whereas the addition of RBCs flattens the classical
Poiseuille parabolic profile, as well documented in the literature. In the `rigid' case (${\rm k}_{\rm ctc} = 10 {\rm k}_{\rm rbc}$),
malignant cells are rapidly pushed out from the center of the capillary and confined to move within
the CFL (\textbf{Fig.\ref{fig5}b}). This indeed increases the likelihood of building adhesive interactions with the
wall. For the other two cases (${\rm k}_{\rm ctc} = 0.5{\rm k}_{\rm rbc}$ and ${\rm k}_{\rm ctc} = {\rm k}_{\rm rbc}$), malignant cells are not observed to marginate
within the considered simulation time. This is due to the fact that RBCs and CTCs
would move similarly in the channel, deforming under flow and moving away from the walls
(\textbf{Fig.\ref{fig5}b,c}).
\\
In \textbf{Fig.\ref{fig5}d}, the CTC trajectories are presented for the three considered cases. In the `soft' and `equal'
cases (red and blue lines), malignant cells migrate towards the centerline $z = H / 2$ and stay within the
capillary core without interacting with the vessel walls throughout the simulation time. Conversely,
in the `rigid' case, malignant cells deviate from the streamlines and, eventually, reach the capillary
wall (margination). For these simulations, the cell-wall adhesion potential is turned off and a moderate
repulsive force is included only to prevent body compenetration.

\subsection{Adhesion dynamics of circulating tumor cells in a large microcapillary: the $25 \mu \rm m$ case}

After marginating towards the vessel walls, CTCs could firmly adhere to the endothelial cells, if proper
conditions are met. The margination and vascular adhesion are fundamental steps in the cascade of
events regulating the extravasation of both leukocytes, in inflammation, and CTCs, in metastasis.
Adhesive interactions are governed by receptor molecules, expressed on the vascular endothelial
cells and ligand molecules, distributed over the CTC membrane (\textbf{Fig.\ref{fig2}}). These molecular
interactions operate only, and only if, CTCs are sufficiently close to the wall, namely closer than a critical
distance set to $d_{\rm wall} = 100 \ \rm nm$. The molecular bonds are computationally treated as linear springs,
whose strength is dictated by the adhesive number $ \rm Ad = \sigma / \nu \dot{\gamma} {\rm r}_{ \rm ctc}$. Simulations
are performed by fixing
a value of the adhesive strenght $\sigma$ and varying the CTC stiffness ${\rm k}_{\rm ctc}$. In this section, the proposed
hierarchical computational model is applied to predict the vascular adhesion dynamics of CTCs as a
function of their deformability.
\begin{figure*}[h!]
 \centering
 \includegraphics[width=15cm]{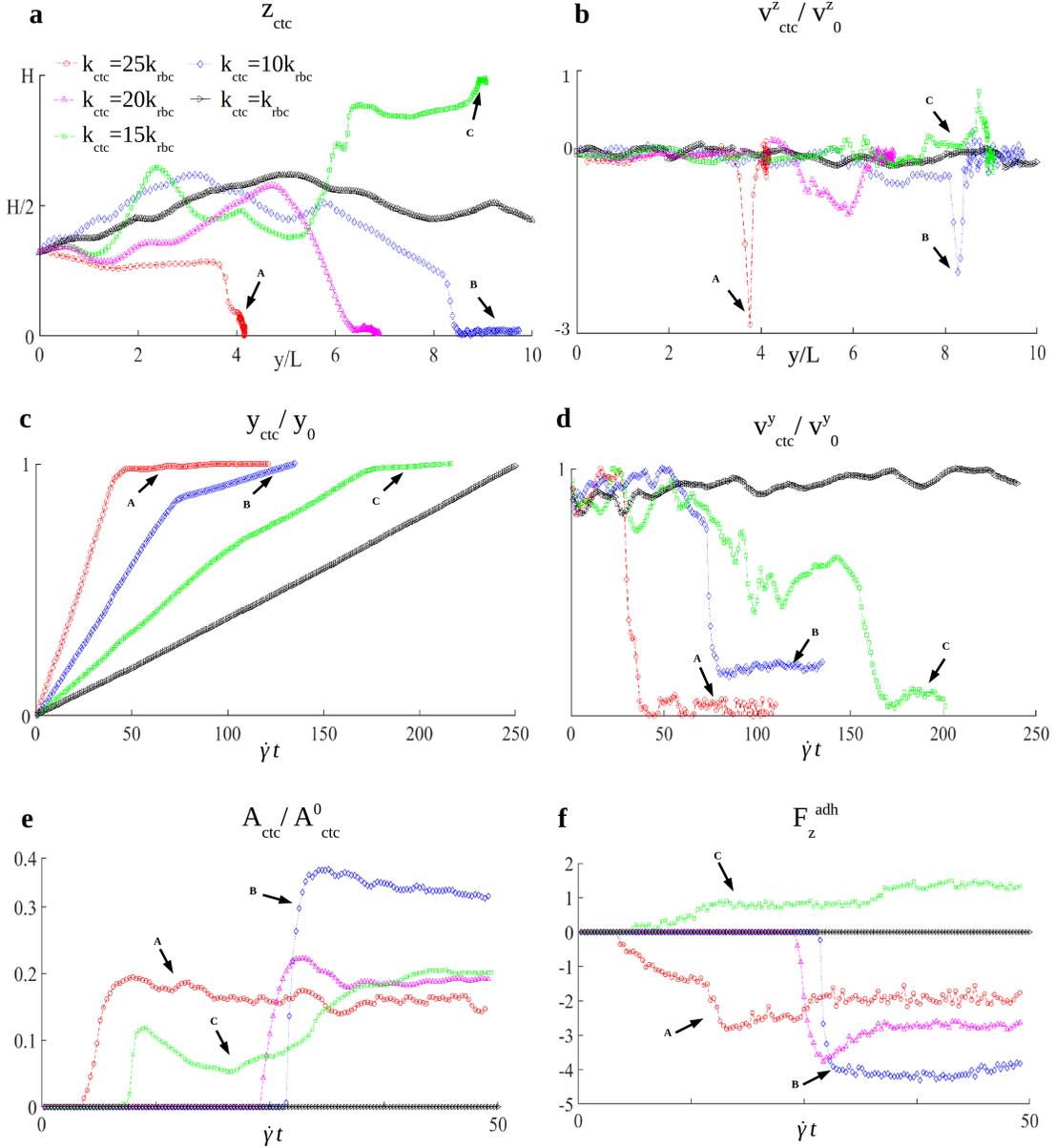}  
 \caption{ \textbf{Adhesion dynamics of a CTC in a large microcapillary ($25 \mu \rm m$ case).} 
 \textbf{a}.Variation over
the capillary length $y/L$ of the vertical position $z_{\rm ctc} / z_0$ of the CTC, for different values of ${\rm k}_{ \rm ctc}$.
\textbf{b}. Variation over the capillary length $y/L$ of the vertical CTC velocity $v^z_{\rm ctc} /v^0_z$, for different
values of ${\rm k}_{\rm ctc}$. \textbf{c}. Variation of the normalized coordinate along the flow $y_{\rm ctc} / y_0$ of the CTC over
time $\dot{\gamma} t$, for different values of ${\rm k}_{\rm ctc}$. \textbf{d}. Variation of the normalized velocity along the flow
$v^y_{\rm ctc} /v^y_0$ of the CTC over time $\dot{\gamma} t$, for different values of ${\rm k}_{\rm ctc}$. \textbf{e}.
Variation of the contact area
$A_{\rm ctc} / A^0_{\rm ctc}$ of the CTC over time $\dot{\gamma} t$. \textbf{f}. Variation of the vertical component of the adhesion
force $F^{\rm adh}_z$
of the CTC over time $\dot{\gamma}t$. 
(black line: ${\rm k}_{\rm  ctc }={\rm k}_{\rm  rbc}$; 
 blue line: ${\rm k}_{\rm  ctc} =10 {\rm k}_{\rm  rbc}$;
green line: ${\rm k}_{\rm  ctc} =15 {\rm k}_{\rm  rbc}$; 
violet line: ${\rm k}_{\rm  ctc }=20 {\rm k}_{\rm  rbc}$;
red line: ${\rm k}_{\rm  ctc} =25 {\rm k}_{\rm  rbc }$).}
 \label{fig6}
\end{figure*}
\\
Based on the results of the previous paragraph, the CTC stiffness is here assumed to be high enough
to allow rapid margination, namely ${\rm k}_{\rm ctc} =10, 15, 20$ and $25{\rm k}_{\rm rbc}$. For soft CTCs,
margination would not occur in capillaries with a tube size of $D=25 \ \mu$m, within the time of the
simulations. Indeed, this implies that soft CTCs could interact with the vascular walls only
in capillaries slightly larger, or even smaller, than the CTC tube size, as shown in the sequel. The
relative position of CTCs, the size of the adhesion area and the adhesion forces are monitored over
time. The CTC dynamics is presented in \textbf{Fig.\ref{fig6}a,b} in terms of the vertical coordinate $z_{\rm ctc}(y)$ and
corresponding velocity component $v^z_{\rm ctc}(y)$ as a function of the position along the $y$-axis, and of the
horizontal coordinate $y_{\rm ctc}(t)$ and corresponding velocity component $v^y_{\rm ctc}(t)$ as a function of time (\textbf{Fig.\ref{fig6}c,d}). The
$z$-direction is normal to the flow, whereas the $y$-direction is aligned with the flow.
\\
In \textbf{Fig.\ref{fig6}a} the black line corresponds to `soft' CTCs (${\rm k}_{\rm ctc} = {\rm k}_{\rm rbc}$) for which margination does not
occur within the considered simulation time. In this case, the vertical position $z_{\rm ctc}$ of the CTC oscillates
sligthly around the centerline of the capillary ($z=H/2$) while the cell is transported downstream. This
is also confirmed by the variation over time of $y_{\rm ctc}$, as shown in \textbf{Fig.\ref{fig6}c}. The black line grows
steadily over time implying that the CTC is steadily moving along the flow direction. As such, the
velocity components $v^z_{\rm ctc}$
along $z$, in \textbf{Fig.\ref{fig6}b} and $v^y_{\rm ctc}$
along $y$ in \textbf{Fig.\ref{fig6}d} are, respectively, nearly
zero and quasi-constant but larger than zero. Correspondingly, the size of the adhesion area $A_{\rm ctc}$ and
adhesion force $F^{\rm adh}_z$ are both null (\textbf{Fig.\ref{fig6}e} and \textbf{f}).
\\
A totally different behavior is documented for `rigid' CTCs (${\rm k}_{\rm ctc} = 25{\rm k}_{\rm rbc}$), as per the red lines in
\textbf{Figs.\ref{fig6}}. The cell is pushed downstream by the flow but also laterally towards the vessel walls until
a stable adhesion is established (point A). This is documented in \textbf{Figs.\ref{fig6}a} and \textbf{c} by the
reduction of $z_{\rm ctc}$ till zero (interaction with the lower vessel wall) and final constant value of $y_{\rm ctc}$.
Also the velocity goes to zero, after a significant spike in $v^z_{\rm ctc}$ due to the margination
process (point A in \textbf{Fig.\ref{fig6}b}). The
area of adhesion $A_{\rm ctc}$ grows upon interaction with the vessel wall and stays quasi-constant over time
(\textbf{Fig.\ref{fig6}e}). Similar observations apply for the adhesion force $F^{\rm adh}$ (\textbf{Fig.\ref{fig6}f}).
This is consistent with
the stable adhesion of a relatively rigid cell that does not squeeze down onto the wall. The case ${\rm k}_{\rm ctc} =
10{\rm k}_{\rm rbc}$ is
depicted in \textbf{Figs.\ref{fig6}} by blue lines. The behavior is quite similar to that of ${\rm k}_{\rm  ctc} = 25{\rm k}_{\rm  rbc}$,
whereby the cell moves downstream and laterally towards the vessel wall and starts interacting with
its surface. However, the margination process occurs over a longer time (see \textbf{Fig.\ref{fig6}a} and \textbf{c}), with
a smoother variation in the velocities (see \textbf{Figs.\ref{fig6}b} and \textbf{d}). Interestingly, and differently from the
more rigid case, the CTC preserves a non-zero $v^y_{\rm ctc}$
velocity, which implies that firm adhesion is not
established but the cell is rather rolling steadly over the vessel wall (point B). Similarly, the size of
the adhesion area and the value of the adhesion force do not change significantly over time after an
initial increase (see \textbf{Figs.\ref{fig6}e} and \textbf{f}). The stable rolling is supported by the continuous rupture and
formation of ligand-receptor bonds, respectively, of the tail and leading edge of the adhesive area.
Finally, the case ${\rm k}_{\rm  ctc} = 15{\rm k}_{\rm  rbc}$ is depicted by the green line, in \textbf{Figs.\ref{fig6}}.
Under this condition, the CTC
exhibits an even slower approach to the vessel wall. Also adhesion does not appear to be firm and
complete suggesting a `crawling' behavior over the vessel wall (point C). The velocity $v^y_{\rm ctc}$
is close to
zero but not null (\textbf{Fig.\ref{fig6}d}), the adhesion area grows steadily with time just like for the adhesion
force documenting a progressive CTC flattening over the wall (see \textbf{Figs.\ref{fig6}e} and \textbf{f}). Indeed, the
higher CTC deformability favors its continuous deformation and conformation to the vessel walls
under hemodynamic forces.
\\
\textbf{Figs.\ref{fig7}} shows the RBC distribution and velocity field around the CTC under firm adhesion
(\textbf{Fig.\ref{fig7}a}) and steady rolling (\textbf{Fig.\ref{fig7}b}) at different time points. \textbf{Fig.\ref{fig7}a}
shows images of the CTC undergoing firm adhesion on the vessel wall, after margination at different
time points (${\rm k}_{\rm  ctc} = 25{\rm k}_{\rm  rbc}$). When the CTC reaches the substrate, the adhesion
force dominates over the lift hydrodynamic force. In this case the cell firmly adheres to the wall and experiences small
variations in configurations due to the complex whole blood flow dynamics. The streamlines of the
velocity $u / u_0$ show the initiation of a local recirculation area around the adhering CTC.
The adhered CTC becomes an obstacle for the red blood cells, which are continuously hitting over
the trailing edge of the CTC, and might detach the cell if adhesion is insufficiently strong. In
\textbf{Fig.\ref{fig7}b}, a CTC undergoing stable rolling dynamics on the wall is shown (${\rm k}_{\rm  ctc} = 10{\rm k}_{\rm  rbc}$). The CTC
moves along the vessel wall with a constant size of the adhesion area at the interface between the
particle and the substrate. The rolling behaviour is confirmed by following the red spot on the cell
membrane over time.
It can be appreciated the increase of the contact area over time which is
associated with the progressive flattening of the CTC over the vessel wall. Although a simplified adhesive model is used based 
on elastic springs, it is capable to predict different adhesive regimes. In particular, it is observed that, if sufficiently soft, 
CTCs can detach from the wall of a capillary. 

\begin{figure}[h!]
 \centering
 \includegraphics[width=15cm]{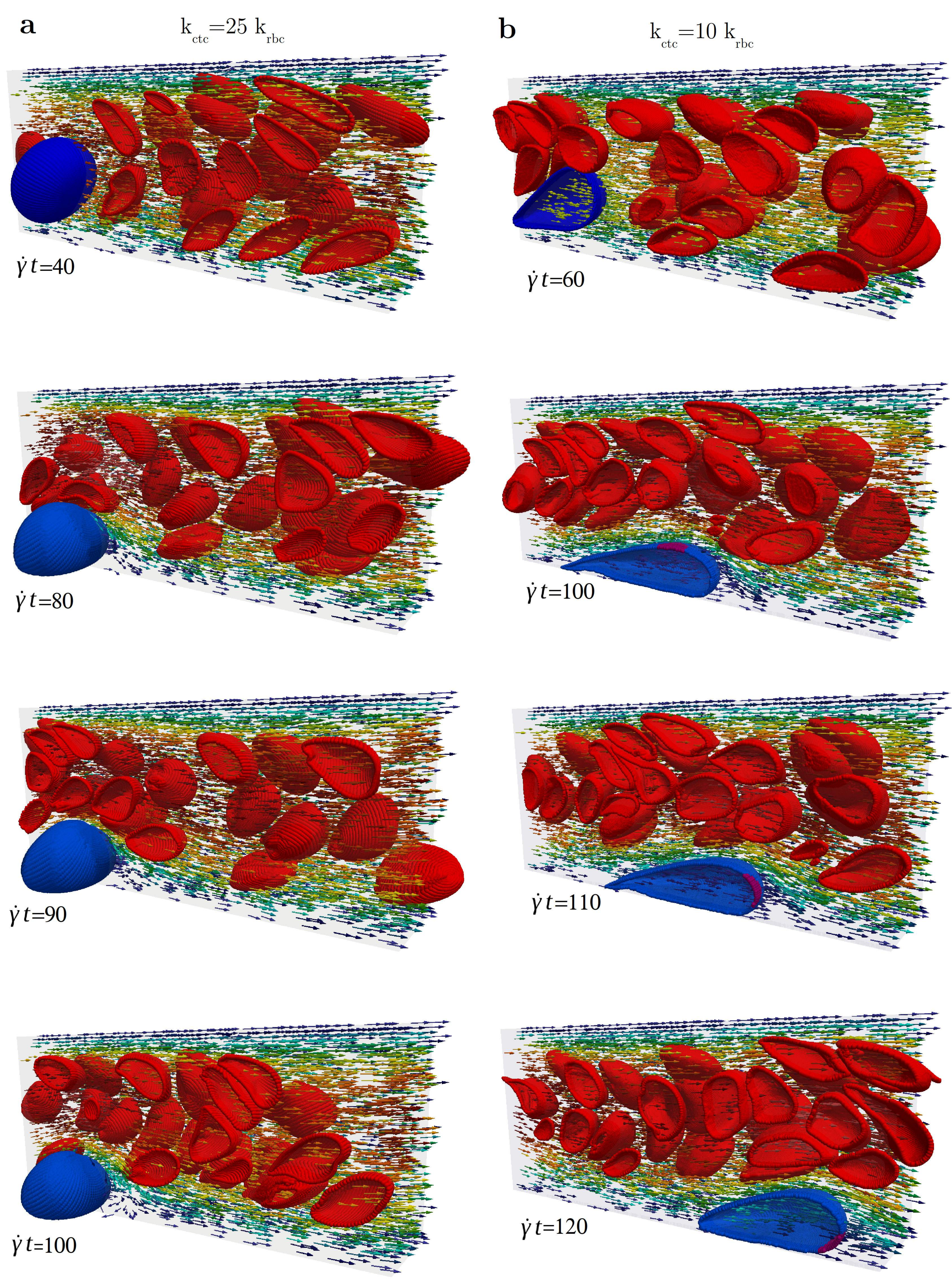}
 \caption{ \textbf{Adhesion dynamics of a CTC in a large microcapillary ($25 \mu \rm m$ case).} 
\textbf{a}. Section of the capillary showing a CTC firmly adhering on the endothelium (${\rm k}_{ \rm ctc}=25 {\rm k}_{\rm rbc}$).
\textbf{b}. Section of the capillary showing a CTC rolling on the bottom of the endothelium (${\rm k}_{\rm ctc}=10 {\rm k}_{\rm rbc}$).
A small portion of the boundary of the cell is labelled in magenta. 
 }
 \label{fig7}
\end{figure}

\subsection{Adhesion dynamics of circulating tumor cells in a small microcapillary: the $15 \mu \rm m$ case}
The
analysis of the margination and adhesion process is here conducted in a capillary having a tube size
comparable with the size of the CTC. The relative position of CTCs, the size of the adhesion area and
the adhesion forces are monitored over time. The CTC dynamics is presented in \textbf{Fig.\ref{fig8}a,b} in terms
of the vertical coordinate $z_{\rm ctc} (y)$ and corresponding velocity component $v^z_{\rm ctc}(y)$ as a function of the
position along the $y$-axis, and of the horizontal coordinate $y_{\rm ctc} (t)$ and corresponding velocity
component $v^y_{\rm ctc}(t)$ as a function of time. The red and blue lines in \textbf{Fig.\ref{fig8}a,b} correspond to `rigid'
CTCs, with ${\rm k}_{\rm ctc} = 25{\rm k}_{\rm rbc}$ and ${\rm k}_{\rm ctc} = 20{\rm k}_{\rm  rbc}$. For these values, the vertical position
$z_{\rm ctc}$ remains close to
the centerline of the capillary ($z=H/2$). Correspondingly, $y_{\rm ctc}$ (red and blue
lines in \textbf{Fig.\ref{fig8}c}) is linearly increasing in time. As such, the velocity components $v^z_{\rm ctc}$
along $z$, in
\textbf{Fig.\ref{fig8}b} and $v^y_{\rm ctc}$
along y in \textbf{Fig.\ref{fig8}d} are, respectively, nearly zero and quasi-constant but larger
than zero. Also, for `rigid' CTCs (${\rm k}_{\rm ctc} = 25{\rm k}_{\rm rbc} , {\rm k}_{\rm ctc} = 20{\rm k}_{\rm rbc}$),
the size of the adhesion area $A_{\rm ctc}$ and
adhesion force $F^{\rm adh}$ are both null (\textbf{Fig.\ref{fig8}e} and \textbf{f}). In the `rigid' case, CTCs are transported along the
flow direction by the fast moving RBCs and do not appear to be interacting with the vessel walls. 
Very different is the behavior observed for a `soft' CTCs. This is shown in \textbf{Fig.\ref{fig8}a} (green line: ${\rm k}_{\rm ctc} =
10{\rm k}_{\rm rbc}$ ; violet line: ${\rm k}_{\rm ctc} = 5{\rm k}_{\rm rbc}$ ; and black line: ${\rm k}_{\rm ctc} = {\rm k}_{\rm rbc}$). 
$z_{\rm ctc}$ increases till the top capillary wall is reached. Spikes in the velocity component $v^z_{\rm ctc}$
(green, violet and black line in \textbf{Fig.\ref{fig8}b})
demonstrate the interaction with the top wall and the subsequent establishment of adhesive
interactions. In \textbf{Fig.\ref{fig8}c} the components in the flow direction $y_{\rm ctc}$ for `soft' CTCs, slightly deviate
from the straight line and the velocity component $v^y_{\rm ctc}$
decreases, as in \textbf{Fig.\ref{fig8}d} (green, violet and
black line). This implies that the cell is no longer in the fluid phase and has established an adhesive
interaction with the vessel walls. In \textbf{Fig.\ref{fig8}e,f}, the variations of the contact area and adhesion
force are shown for `soft' CTCs (green, violet
and black line). In all the cases, the contact area and the adhesion force increases with time.
This is due to the interaction with RBCs and hydrodynamic
lift force, which affects the formation of the contact area.
\begin{figure*}[h!]
 \centering
 \includegraphics[width=15cm]{fig8-eps-converted-to.pdf}
 \caption{ \textbf{Adhesion dynamics of a CTC in a small microcapillary ($15 \mu \rm m$ case).} 
 \textbf{a}.Variation over
the capillary length $y/L$ of the vertical position $z_{\rm ctc} / z_0$ of the CTC, for different values of ${\rm k}_{ \rm ctc}$.
\textbf{b}. Variation over the capillary length $y/L$ of the vertical CTC velocity $v^z_{\rm ctc} /v^z_0$, for different
values of ${\rm k}_{\rm ctc}$. \textbf{c}. Variation of the normalized coordinate along the flow $y_{\rm ctc} / y_0$ of the CTC over
time $\dot{\gamma} t$, for different values of ${\rm k}_{\rm ctc}$. \textbf{d}. Variation of the normalized velocity along the flow
$v^y_{\rm ctc} /v^y_0$ of the CTC over time $\dot{\gamma} t$, for different values of ${\rm k}_{\rm ctc}$. \textbf{e}.
Variation of the contact area
$A_{\rm ctc} / A^0_{\rm ctc}$ of the CTC over time $\dot{\gamma} t$. \textbf{f}. Variation of the vertical component of the adhesion
force $F^{\rm adh}_z$
of the CTC over time $\dot{\gamma}t$. (
black line: ${\rm k}_{\rm  ctc }={\rm k}_{\rm  rbc}$;
red line: ${\rm k}_{\rm  ctc} =25 {\rm k}_{\rm  rbc }$;
blue line: ${\rm k}_{\rm  ctc} =20 {\rm k}_{\rm  rbc}$;
green line: ${\rm k}_{\rm  ctc} =10 {\rm k}_{\rm  rbc}$;
violet line: ${\rm k}_{\rm  ctc }=5 {\rm k}_{\rm  rbc}$).
 }
 \label{fig8}
 \end{figure*}
\\
\textbf{Fig\ref{fig9}.a} shows images of a CTC undergoing `train' dynamics in a small capillary ($D=15 \ \mu$m) , at
different time points for ${\rm k}_{\rm ctc} = 25{\rm k}_{\rm rbc}$. The `rigid' CTC is transported downstream without marginating.
This results in a stable dynamics in which the CTC becomes a moving obstacle inside the capillary.
The RBCs pile up behind the cell and form a dense aggregate \cite{Takeishi2015}. \textbf{Fig\ref{fig9}.b} shows
representative images of the dynamics of a `soft' CTC (${\rm k}_{\rm ctc} = {\rm k}_{\rm rbc}$) in a small capillary. Due to its
deformability the cell interacts with the vessel walls establishing adhesion. This results in a partial
occlusion of the capillary. RBCs tend to pile up behind the cell, continuously pushing the
cell against the wall, until they are free to pass again.
\\
It can be concluded that in small capillaries of size comparable with that of the CTC, the stiffness of
the cell is responsible for a transition from a `train' dynamics, in which the CTC moves along the
capillary without interacting with the walls (`rigid' case), to margination and subsequent adhesion dynamics (`soft' case).

\begin{figure}[h!]
 \centering
 \includegraphics[width=15.0cm]{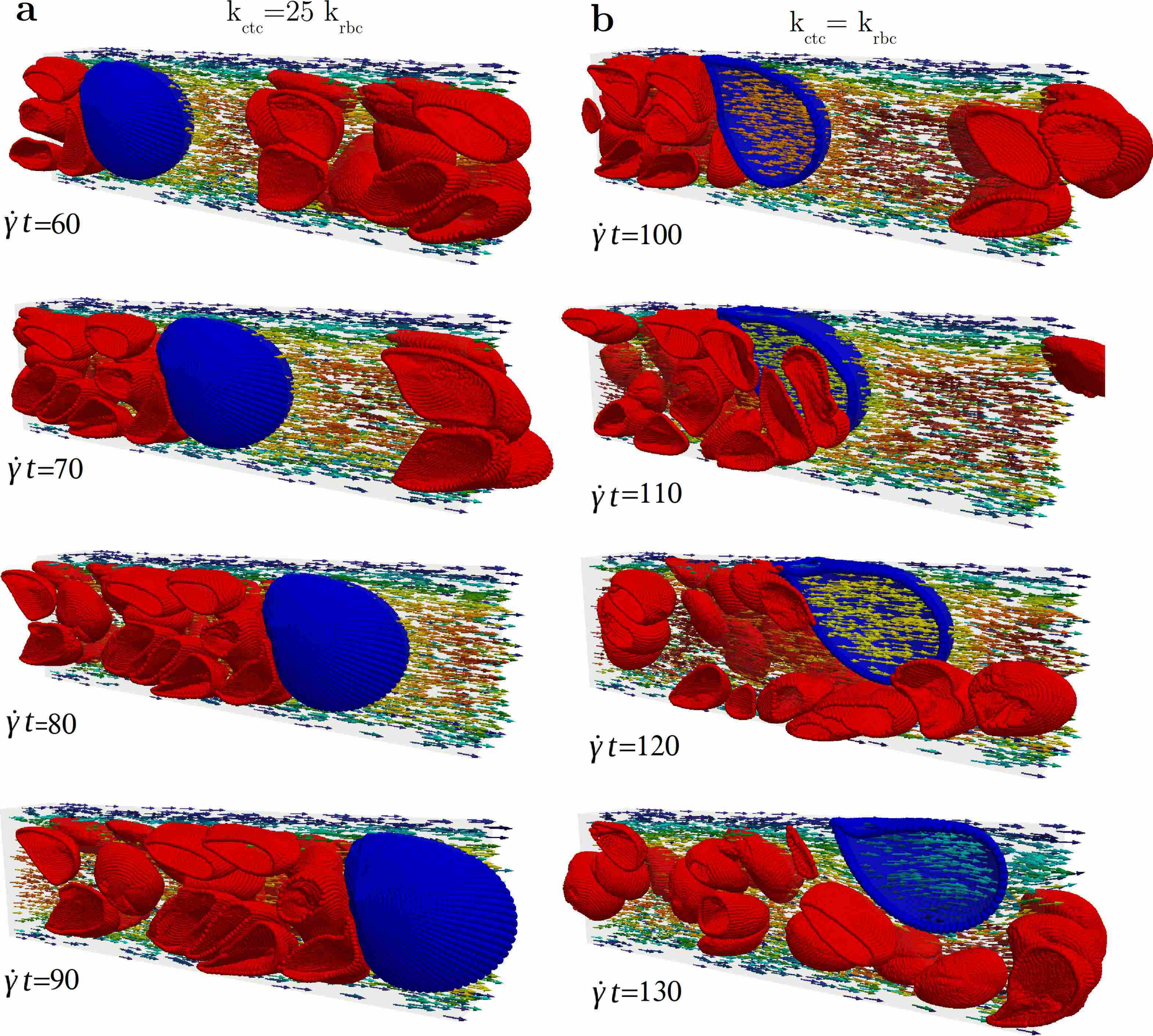}
 \caption{ \textbf{Adhesion dynamics of a CTC in a small microcapillary ($15 \mu \rm m$ case).}
\textbf{a}. Section of the channel showing the so called  `train' dynamics of a CTC (${\rm k}_{\rm ctc}=25 {\rm k}_{\rm rbc}$).
\textbf{b}. Section of the capillary showing a CTC progressively adhering on the top of the capillary (${\rm k}_{\rm ctc}={\rm k}_{\rm rbc}$).
 }
 \label{fig9}
\end{figure}

\section{Conclusions}
Using a combined Lattice Boltzmann-Immersed Boundary method, the transport of deformable CTCs in a whole blood
capillary flow has been analyzed in terms of cell displacements and velocities, and cell interactions with the
vessel walls. The evolution over time of the area of adhesion and adhesion forces exchanged at the cell-wall
interface has been documented for CTC stiffer, softer and equally deformable as compared to RBCs. It has been
demonstrated that the interaction between deformable CTCs and RBCs is crucial in shaping the metastatic process.
\\
Rigid CTCs have been observed to marginate rapidly within $25 \ \mu$m microcapillaries and efficiently interact with
the vessel walls as they are pushed laterally by the RBCs. On the contrary, in smaller $15 \ \mu$m microcapillaries,
rigid CTCs cannot establish firm interactions with the vessel walls. This should be ascribed to the significant
flow obstruction induced by a rigid CTC adhering in such a small vessel. In other words, the fast moving RBCs
in $15 \ \mu$m microcapillaries form a compact train that constantly pushes and dislodge downstream any obstacle,
such as a rigid CTC. Different adhesive regimes have been predicted for the rigid CTCs depending on their relative
stiffness to RBCs. Very rigid CTC would firmly adhere, if proper local hemodynamic and biophysical conditions ar met.
Intermediate rigid CTCs would roll over the vascular walls, whereas CTCs that are slightly stiffer than RBCs could
crawl over the surface as a combination of rolling and progressive squeezing against the wall.
Soft CTCs have been observed to deform and navigate together with the RBCs in the core of the blood vessel in
$25 \ \mu$m microcapillaries. Very differently, in smaller $15 \ \mu$m microcapillaries, soft CTCs can deform and squeeze
progressively within the train of fast moving RBCs and the vessel walls. This indeed increases the surface area
of the CTC exposed to the vessel wall and inevitably favors firm adhesion. As a consequence, soft CTCs are expected
to have a higher longevity in blood and, possibly, the ability to evade more efficiently than rigid CTC the
recognition by cells of the immune system.
These findings highlight the role of CTC deformability in defining the metastatic potential of cancer cells.
\section{Acknowledgements}
This project was partially supported by the European Research Council, under the European
Union's Seventh Framework Programme (FP7/2007-2013)/ERC grant agreement no. 616695,
by the Italian Association for Cancer Research (AIRC) under the individual investigator grant
no. 17664, and by the European Union's Horizon 2020 research and innovation programme under
the Marie Sk{\l}odowska-Curie grant agreement no. 754490.
\section{Conflicts of interest}
Dr. Lenarda, Dr. Coclite and Dr. Decuzzi declare that they have no conflicts of interest.
\section{Ethical standards}
No animal studies or experiments with human samples were carried out by the authors for this article.


\end{document}